\documentclass[namedreferences,hyperref,optionalrh,solaromanenum]{spr-sola}

\usepackage{graphicx}                    
\usepackage{color}                       

\newcommand{\aap}{{\it Astron. Astrophys.}}

\newcommand{\apj}{{\it Astrophys. J.}}
\newcommand{\apjl}{{\it Astrophys. J. Lett.}}
\newcommand{\apjs}{{\it Astrophys. J. Supp.}}
\newcommand{\apss}{{\it Astrophys. Space Sci.}}

\newcommand{\grl}{{\it Geophys. Res. Lett.}}

\newcommand{\mnras}{{\it Mon. Not. R. Astron. Soc.}}

\newcommand{\pasa}{{\it Publ. Astron. Soc. Aus.}}

\newcommand{\solphys}{{\it Sol. Phys.}}
 
\newcommand{\ssr}{{\it Space Sci. Rev.}} 
\newcommand{\zap}{{\it Zeitschrift für Astrophysik}} 
\chardef\us=`\_

\begin{document}

\begin{frontmatter}

\title{Solar Filament Physiognomy: Inferring Magnetic Quantities from Imaging Observations}

\author[addressref={aff1,aff2,aff3},email={chenpf@nju.edu.cn}]{\inits{P.F.}\fnm{P.~F.}~\snm{Chen}\orcid{0000-0002-7289-642X}}

\address[id=aff1]{School of Astronomy and Space Science, Nanjing University, Nanjing 210023, PR China}
\address[id=aff2]{Key Laboratory for Modern Astronomy and Astrophysics, Nanjing University, Nanjing 210023, PR China}
\address[id=aff3]{State Key Laboratory of Lunar and Planetary Sciences, Macau University of Science and Technology, Macau 999078, PR China}

%
\runningauthor{Chen}
\runningtitle{Solar Filament Physiognomy}


\begin{abstract}
Magnetic field is the key physical quantity in solar physics as it controls all kinds of solar activity, ranging from nanoflares to big flares and coronal mass ejections (CMEs). However, so far only the magnetic field on the solar surface can be more or less precisely measured, and the most important coronal magnetic field remains undetectable accurately. Without the knowledge of the coronal magnetic field, it is even more difficult to obtain secondary quantities related to magnetic field, such as the magnetic helicity and magnetic configuration, including the curvature of field lines. The prevailing approaches to obtain the coronal magnetic field include coronal magnetic extrapolation and coronal seismology. Actually there were scattered efforts to derive secondary magnetic quantities based on imaging observations of solar filaments, without the help of polarization measurements. We call this approach solar filament physiognomy. In this paper, we review these efforts made in the past decades, and point out that this approach will be promising as large telescopes are being built and more fine structures of filament channels will be revealed.
\end{abstract}

%

\end{frontmatter}

%
\section{Introduction}\label{s:1} 

Solar filaments are cold dense plasmas embedded in the hot tenuous corona. They appear as elongated dark structures on the solar disk in H$\alpha$ or extreme ultraviolet (EUV) images. When they follow the solar rotation to the solar limb, they are seen to be suspended above the solar surface in the corona, hence are also called prominences. High-resolution observations revealed that a filament is composed of a bundle of thin threads, with a length of 5--20 arcsec and a width of 0.3 arcsec \citep{liny05}. Solar filaments are intriguing in several senses. First, many filaments are formed through coronal condensation. Hence, their formation can be considered as the inverse process of coronal heating. It is argued that investigating filament formation might shed unique light on understanding the mechanism of coronal heating, one of the most important problems in astrophysics \citep{chen20}. Second, solar filaments are often the source region of solar flares and coronal mass ejections \citep[CMEs,][]{chen11, parenti14, zou19}. In particular, filament eruptions, solar flares, and CMEs are different aspects of the same process in the standard flare/CME model. Several review papers about solar filaments have been published, covering spectral diagnosis of their thermal structure \citep{labr10}, their magnetic structure \citep{mack10}, their observational features \citep{parenti14}, and their magnetic models \citep{gibs18}. In this short review paper, we pick up a specific topic on how to infer the magnetic structure of solar filaments based on their observational appearance, which was not emphasized in the previous reviews.

Besides their thermal structure, the more important property of solar filaments is their magnetic field, including the magnetic strength, distribution, and topology, which determines whether a filament can be formed and when it might erupt. However, the magnetic field in the solar corona is hard to be precisely measured, and only the vector magnetograms on the solar surface are routinely recorded. To obtain the magnetic information in the solar corona, we mostly rely on the coronal magnetic field extrapolation based on the potential or force-free models with the photospheric magnetograms as the bottom boundary conditions. Unfortunately, such an extrapolation is an ill-posed problem, and different methods result in different magnetic distribution in the corona. More importantly, the extrapolation with the existing methods may inevitably miss some singular magnetic topology which is crucial for the triggering and eruption of solar flares and CMEs. Alternatively, colleagues tried to infer the coronal magnetic field based on coronal loop oscillations, filament oscillations, and various waves. This approach is called coronal seismology, which is still improving as the error bar is not small enough and it is still difficult to derive the magnetic distribution of the source region hosting solar eruptions. 

\begin{figure} 
	\centerline{\includegraphics[width=1.0\textwidth,clip=]{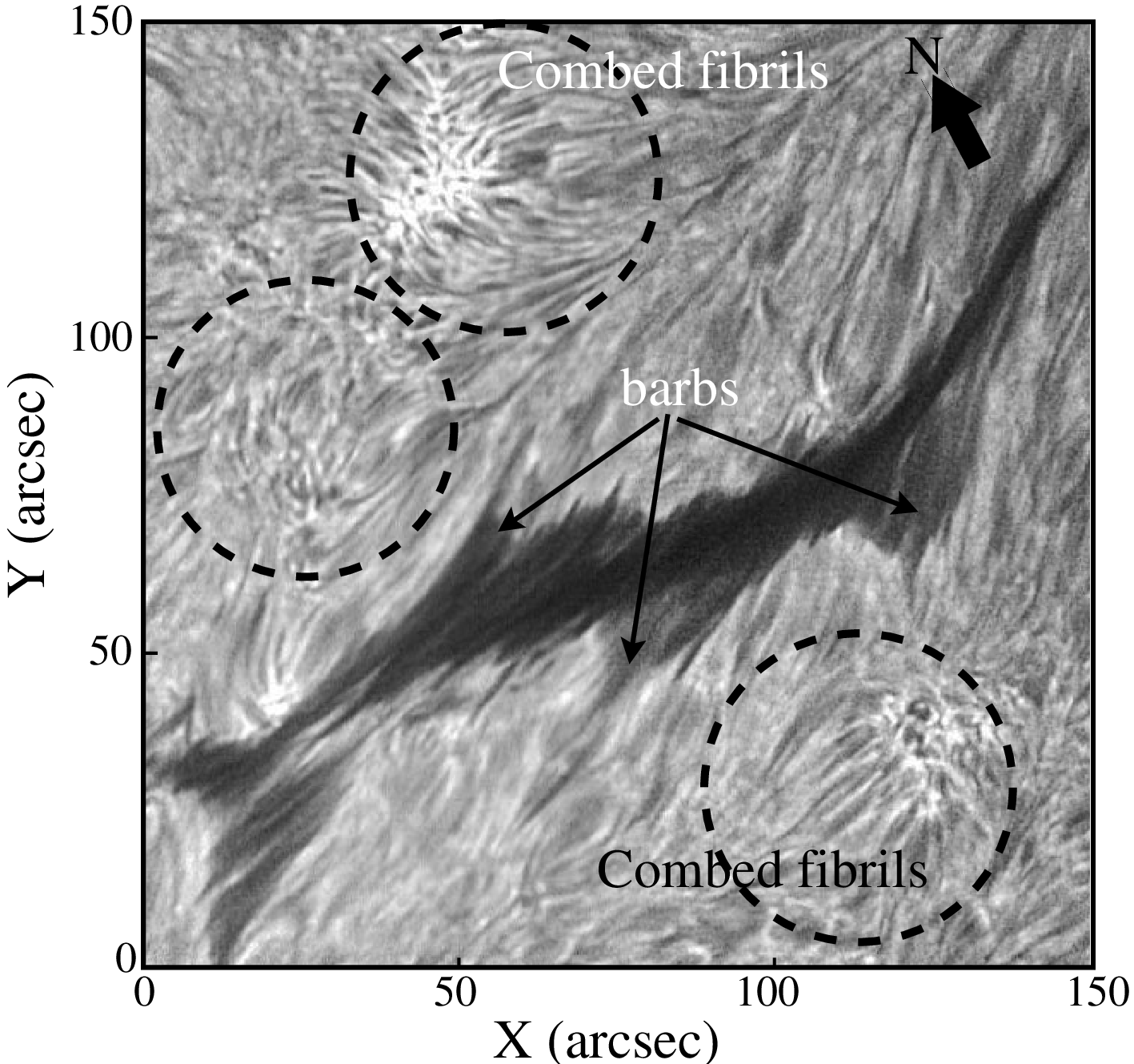}}
	\caption{H$\alpha$ image of a quiescent filament with several barbs inside a filament channel observed by the New Vacuum Solar Telescope \citep{liuz14} on 2025 April 26. A collection of threads form the elongated filament spine with several barbs. The northern direction is indicated by the thick arrow. The combed dark fibrils originating from the bright plagettes are mainly directed toward right in the northern side of the filament and are directed toward left in the southern side. }\label{fig01}
\end{figure}

Efforts have been made to measure the magnetic field of solar filaments directly \citep{rust67, lero78, bomm94}. This kind of measurements are based on Zeeman or Hanle effects of He {\small I} D3 or He {\small I} 10830 \AA\ lines. However, it it noticed that a solar filament occupies only small portion of the filament channel \citep{gibs15, guojh22}. Specifically, solar filaments are located around the magnetic dips of twisted or sheared magnetic field lines, which cannot represent the most intriguing portion of the whole magnetic structure of filament channels. A filament with several barbs inside a filament channel is illustrated in Figure \ref{fig01}, where the filament is the elongated dark structure, and the filament channel corresponds to the region bounded by the combed H$\alpha$ fibrils marked by dashed circles on the two sides of the filament.

%

There is a series of quantities that are used to describe the properties of magnetic field, for example, vector magnetic field $\mathbfit{B}$, electric current density $\mathbfit{J}$, current helicity density $h_c=\mathbfit{J}\cdot \mathbfit{B}$, magnetic twist $T_w$, etc.  Note that the integrated $h_c$ is not conservative, and the integrated magnetic helicity density $H_m=\int \mathbfit{B}\cdot \mathbfit{A} {\rm d}V$ is a conservative quantity, where $\mathbfit{B}=\nabla \times \mathbfit{A}$ with $\mathbfit{A}$ being the magnetic vector potential in Coulomb gauge. Among these quantities, the vector magnetic field $\mathbfit{B}$ is the primary quantity, i.e., all other quantities can only be obtained after we get the 3-dimensional (3D) distribution of $\mathbfit{B}$ in the corona.

Generally speaking, primary quantities are measured with a certain effect in astronomy. For example, plasma velocity is measured with Doppler effect. Similarly, the magnetic field $\mathbfit{B}$ is measured with Zeeman or Hanle effects. However, as mentioned above, we can only measure the vector magnetic field in the photosphere routinely (and the longitudinal magnetic field of solar prominences occasionally). From the photospheric magnetograms to the coronal magnetic field, several steps would bring errors. First, the precision of the horizontal magnetic field in the photosphere is significantly lower than that of the longitudinal component. Second, there exists $180^\circ$ ambiguity in determining the orientation of the horizontal magnetic component. Third, the often-assumed force-free condition is actually not satisfied in the photosphere. Last but not the least, as mentioned above, it is an ill-posed mathematical problem to extrapolate the magnetic field from the solar surface to the corona. Summing up, it is expected to have large errors in those secondary magnetic quantities on top of the sophisticated long procedure. With all these difficulties, the whole community have been thinking for a long time about whether we can precisely determine some properties of those secondary magnetic quantities based on ordinary imaging observations of filament channels. 

In fact, with the efforts of the community in the past decades, we can indeed infer some properties of the secondary magnetic quantities, such as the sign of helicity, the curvature radius of the magnetic field lines, the magnetic configuration, and even the relative strength of magnetic twist in the core region of filament channels. We call this approach solar filament physiognomy, i.e., to infer the magnetic properties from imaging observations without the help of polarization measurements. In this review paper, we summarize the achievements made in the past decades, with each chapter devoted to one magnetic property. It is noted that prominence seismology can also be considered as a part of solar filament physiognomy. However, it is applied to derive the primary quantity, magnetic field, and there are already systematic works on the topic, therefore we neglect this part, and refer the reader to the relevant reviews \citep{arre18}.

\section{Helicity Sign of Filament Channels}\label{s:2} 

Magnetic helicity and current helicity are two of the most important quantities that describe the complexity of nonpotential magnetic field, as they quantify the twist, writhe, and linkage of magnetic field lines in the solar atmosphere \citep{berg84}. First, magnetic helicity characterizes the accumulation of magnetic free energy of the solar eruption hosting regions as nonpotential magnetic field is either strongly sheared or even twisted. Second, the triggering of solar eruptions is related to either magnetoydrodynamic (MHD) instabilities when probably magnetic helicity reaches a critical parameter regime \citep{zhangm06} or magnetic reconnection when magnetic connectivity is changed \citep{chen11}. Third, the sign of magnetic helicity is one key factor that determines how CMEs rotate during propagation from the corona to the interplanetary space \citep{zhouzj22}, hence it is intimately related to the geoeffectiveness of solar eruptions \citep{jing04}.

The helicity of magnetic field can be either described by magnetic helicity $H_m=\int \mathbfit{B}\cdot \mathbfit{A} {\rm d}V$ or current helicity $H_c=\int \mathbfit{J}\cdot \mathbfit{B} {\rm d}V$, where the integral is conducted over a volume $V$. Although $H_m$ is a conservative quantity but $H_c$ is not and it is not necessary for them to have the same sign in general, they do have the same sign in the typical flux rope configurations important to coronal processes \citep{vand03, russ19}. Owing to its simplicity, $H_c$ has been adopted more frequently to investigate the helicity pattern of solar active regions. One intriguing result related to helicity is that there exists a hemispheric preference, i.e., magnetic structures at various scales possess primarily negative helicity in the northern hemisphere and positive helicity in the southern hemisphere, with a preference strength of 60--84\% based on different helicity proxies \citep[see][for a review]{pevt14}.

Strictly speaking, to deduce the helicity sign, one has to measure the vector magnetic field at two layers in the solar lower atmosphere. However, routine observations can provide the vector magnetograms in the photosphere only. Even for the one-layer vector magnetograms, there exists the $180^\circ$ ambiguity in measuring the horizontal magnetic component, which would result in the misidentification of the helicity sign. Fortunately, several alternative methods have been proposed to determine the helicity sign indirectly based on structural patterns of filament channels, without the need of photospheric magnetograms.

\subsection{Fibril Orientation}\label{fibril}

\begin{figure} 
 \centerline{\includegraphics[width=1.0\textwidth,clip=]{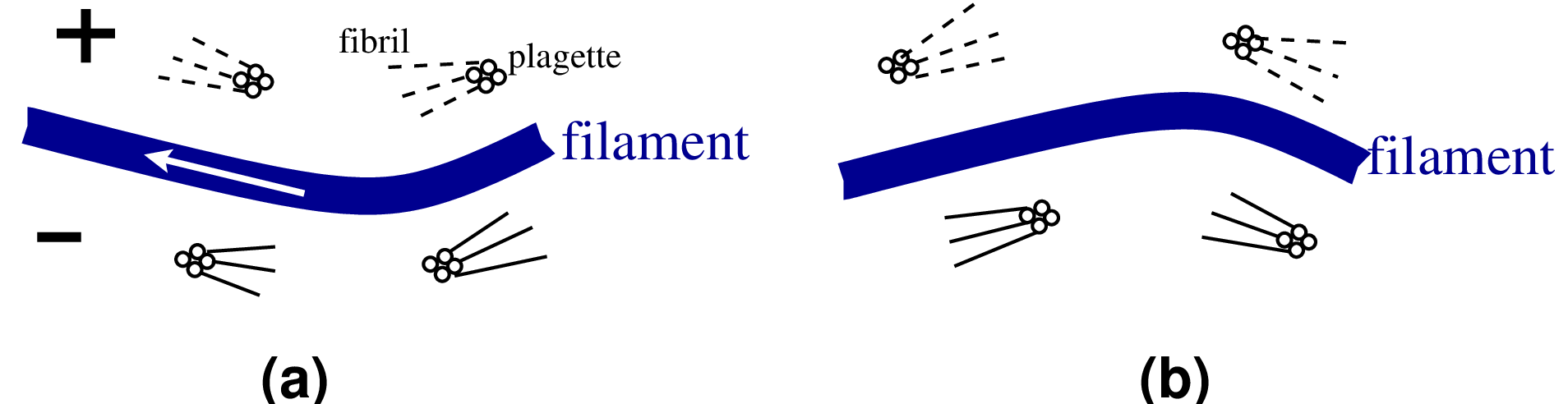}}
 \caption{(a) A method to determine the axial magnetic field and the helicity sign by combining H$\alpha$ filtergrams and longitudinal magnetograms proposed by \citet{fouk71}. (b) Sketch showing how the helicity sign can be determined by H$\alpha$ filtergrams only.}\label{fig02}
 \end{figure}

In the quiet Sun, H$\alpha$ fibrils rooted on bright plagettes are oriented randomly, following the canopy-shaped magnetic configuration near the chromospheric height. However, the H$\alpha$ fibrils inside filament channels are nearly parallel to the filament spine, emanating from bright plagettes toward one direction on one side of the filament channel and toward the other direction on the other side of the filament channel. As shown in Figure \ref{fig01}, the H$\alpha$ fibrils on the two sides of the filament look like being combed. Considering that H$\alpha$ fibrils represent the direction of the chromospheric magnetic field, \citet{fouk71} proposed that one can easily determine the axial component of the magnetic field in the filament and inside the filament channel at large with the information of positive and negative polarities from longitudinal magnetograms. As illustrated in Figure \ref{fig02}(a), the northern side of the filament channel has positive polarity and the southern side of the filament channel has negative polarity, hence the antiparallel fibrils in Figure \ref{fig02}(a) imply that the axial magnetic component of the filament channel is leftward, as indicated by the white arrow. 

%

Once we know the magnetic polarities of the filament channel and the axial magnetic component, we can easily determine the handedness or the heilcity sign \citep{rust94}. Taking Figure \ref{fig02}(a) as an example, the magnetic orientation is consistent with left handedness, hence the helicity is negative. It should be pointed out that \citet{mart94} defined the filament chirality in another set of terminology, i.e., dextral or sinistral. According to them, when viewed from the positive magnetic polarity side of a filament, the chirality of the filament is called dextral when the axial magnetic component is toward right, and is called sinistral when the axial magnetic component is toward left. According to this definition, the filament with left-handedness helicity in Figure \ref{fig02}(a) would also be classified as dextral chirality. Since ``dextral" has the meaning of ``right", such contrast often led to confusion, in particular for newcomers. Therefore, we emphasize here that left-handedness helicity, or negative helicity, corresponds to dextral chirality, and right--handedness helicity, or positive helicity, corresponds to sinistral chirality due to historic reasons. There would be no confusion if \citet{mart94} had happened to determine the filament chirality by the direction of the axial magnetic field viewed from the NEGATIVE magnetic polarity side.

In the original papers of \citet{fouk71} and \citet{mart92}, one needs both longitudinal magnetograms and H$\alpha$ filtergrams together to determine the direction of the axial magnetic component, which can then be used to deduce the helicity sign. However, here in this paper we point out that there is no need of magnetograms if one wants to determine the helicity sign only. We can draw an analogy between the combed fibrils and the bent grass on the ground blown by hurricanes: If the H$\alpha$ fibrils on the two sides of a filament channel look like being blown by a counterclockwise hurricane, as illustrated in Figure \ref{fig02}(a), the helicity should be negative (no matter which side of the filament channel has positive polarity); If the H$\alpha$ fibrils on the two sides of a filament channel look like being blown by a clockwise hurricane, as illustrated in Figure \ref{fig02}(b), the helicity should be positive. According to this rule, the filament in Figure \ref{fig01} has positive helicity as the combed H$\alpha$ fibrils look like being blown by a clockwise hurricane.

\subsection{Orientation of Coronal Cells}\label{cell}

\begin{figure} 
	\centerline{\includegraphics[width=0.8\textwidth,clip=]{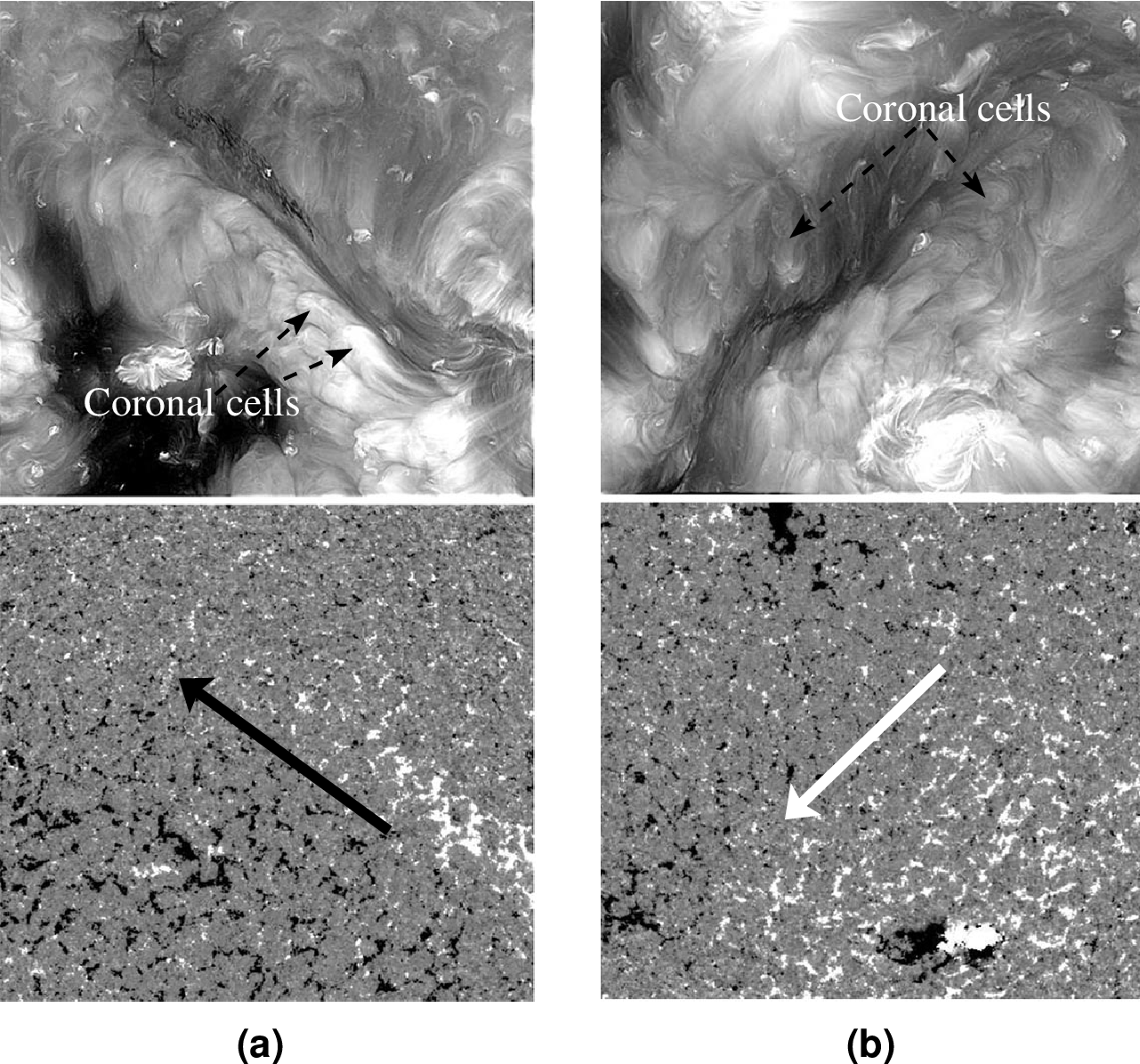}}
	\caption{Examples showing how to determine the axial magnetic field and the helicity sign based on EUV images (top row) and longitudinal magnetograms (bottom row). Adapted from \citet{shee13}. The dashed arrows in the top row indicate four coronal cells as examples, and the solid arrows in the bottom row indicate the deduced direction of the axial magnetic field of the filament channel.}\label{fig03}
\end{figure}

It was shown that cellular features appear in the EUV images with the formation temperature about 1.2 MK, such as the 193 \AA\ waveband \citep{shee12}. As shown in Figure \ref{fig03}, a coronal cell has a round brighter root and branching tapered plumes. In the quiet Sun, they are believed to extend radially, with diameters about 30 Mm and centered on networks. However, along filament channels, the plumes of coronal cells bend horizontally, with the tapered ends pointing toward one direction roughly parallel to the filament spine on one side of the filament channel and toward the opposite direction on the other side of the filament channel. We tend to believe that coronal cells are the coronal extension of H$\alpha$ fibrils.

\citet{shee13} proposed to infer the direction of the axial magnetic field and the helicity of filament channels based on the bending orientation of the coronal cells, similar to the procedure mentioned in \citet{fouk71} and \citet{mart92} using H$\alpha$ fibrils. That is to say, in order to know the direction of the axial magnetic field, we need to combine the EUV images and longitudinal magnetograms. As shown in Figure \ref{fig03}(a), the northern side of the filament channel has positive polarity and its coronal cells are bent to the left, hence the axial magnetic field is toward the top-left, as indicated by the black arrow. Therefore, the helicity of the filament channel is left-handed or negative. As a contrast, in Figure \ref{fig03}(b), the southern side of the filament channel has positive polarity, and its coronal cells bend to the left, hence the axial magnetic field is toward the bottom-left, as indicated by the white arrow. Therefore, the helicity of the filament channel is right-handed or positive. 

Again, as \citet{shee13} mentioned, if we simply want to know the helicity sign, only EUV images are sufficient. They proposed a simple way to determine the helicity sign: When you look at the filament from one endpoint, if the cellular plumes on your left side point toward you, the helicity should be negative, as in the case of Figure \ref{fig03}(a); if the cellular plumes on your left side point away from you, the helicity should be positive, as in the case of Figure \ref{fig03}(b). Here we propose an alternative way: Similar to what was mentioned in Section \ref{fibril}, if the cellular plumes on the two sides of the filament channel bend as like being blown by a counterclockwise hurricane, the filament channel has negative helicity as shown in the top-left panel of Figure \ref{fig03}; if the cellular plumes look like being blown by a clockwise hurricane, the filament channel has positive helicity as shown in the top-right panel of Figure \ref{fig03}.

\subsection{Relative skew of filament threads}\label{thread}

With high-resolution observations, it has been revealed that a solar filament is composed of many dynamic thin threads, with a filling factor about 0.3 or more \citep{kuce98}, which are reproduced later via MHD simulations by \citet{zhou20} and \citet{jerc23}, who proposed that the formation of multiple threads in a filament is due to the turbulent heating of the solar lower atmosphere. It is believed that the filament threads trace the local magnetic field \citep{liny05, hana17}, and occupy different layers of dipped magnetic field \citep{guojh22}. Owing to various reasons, such as inhomogeneity, transient brightenings, and mass flows, we can identify a higher thread as the foreground and a lower thread as the background when a filament is viewed from top. Frequently they are oriented differentially.

\begin{figure} 
	\centerline{\includegraphics[width=1.0\textwidth,clip=]{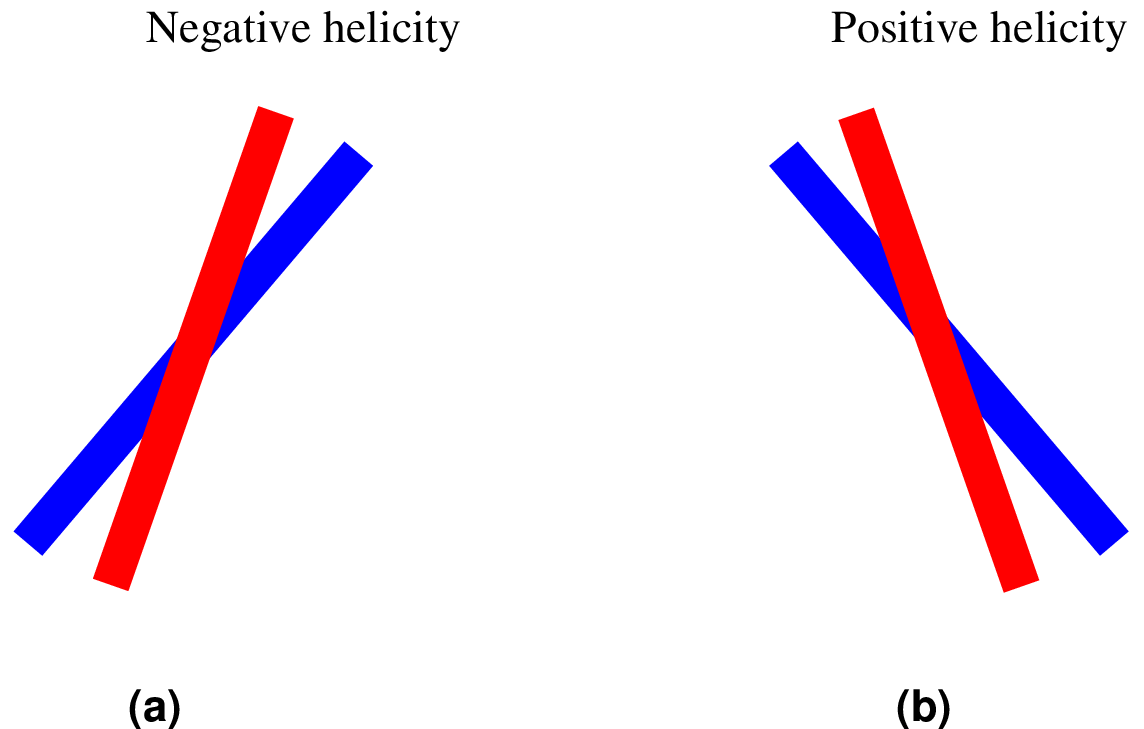}}
	\caption{Sketch showing how to determine the helicity sign based on the orientation of the higher thread ({\it red line}) relative to the lower thread ({\it blue line}) in a filament, a method proposed by \citet{chae00}.}\label{fig04}
\end{figure}

Based on the differential orientations of the lower and higher threads of filaments, \citet{chae00} proposed a method to determine the helicity sign of filaments. His method is illustrated in Figure \ref{fig04}: Relative to the lower thread ({\it blue line}), if the higher thread ({\it red line}) is skewed counterclockwise, the helicity is negative (left panel of Figure \ref{fig04}); If the higher thread ({\it red line}) is skewed clockwise, the helicity is positive (right panel of Figure \ref{fig04}). Application of this method to several filaments confirmed the validity of this method \citep{chae00, chen14}.

\subsection{Handedness of sigmoids}\label{sigmoid}

Solar filaments reside in the core magnetic field of filament channels, which is characterized by strong shear and/or twisted field lines. Further away from the magnetic polarity inversion line, the core field transits to less sheared envelope magnetic field and then to the weakly sheared background magnetic field. As a result, the inner edge of the envelope field often corresponds to a magnetic quasi-separatrix layer (QSL), where magnetic connectivity changes rapidly \citep{demo06}. Therefore, in contrast to the low temperature filament, the envelope field is filled with much hotter and denser plasma than the background corona. As the consequence, the hot dense envelope part of the filament channel looks as twisted bright features in the soft X-ray or EUV images \citep{canf99}, which are called sigmoids.
\citet{naka71} tried to associated S-shaped structures to positive helicity, and inverse-shaped structures to negative helicity, which was confirmed by \citet{rustk96} and \citet{low03} among others. This association provides another method to infer the helicity sign of sigmoids and the embedded filament channel. 

\begin{figure} 
	\centerline{\includegraphics[width=1.0\textwidth,clip=]{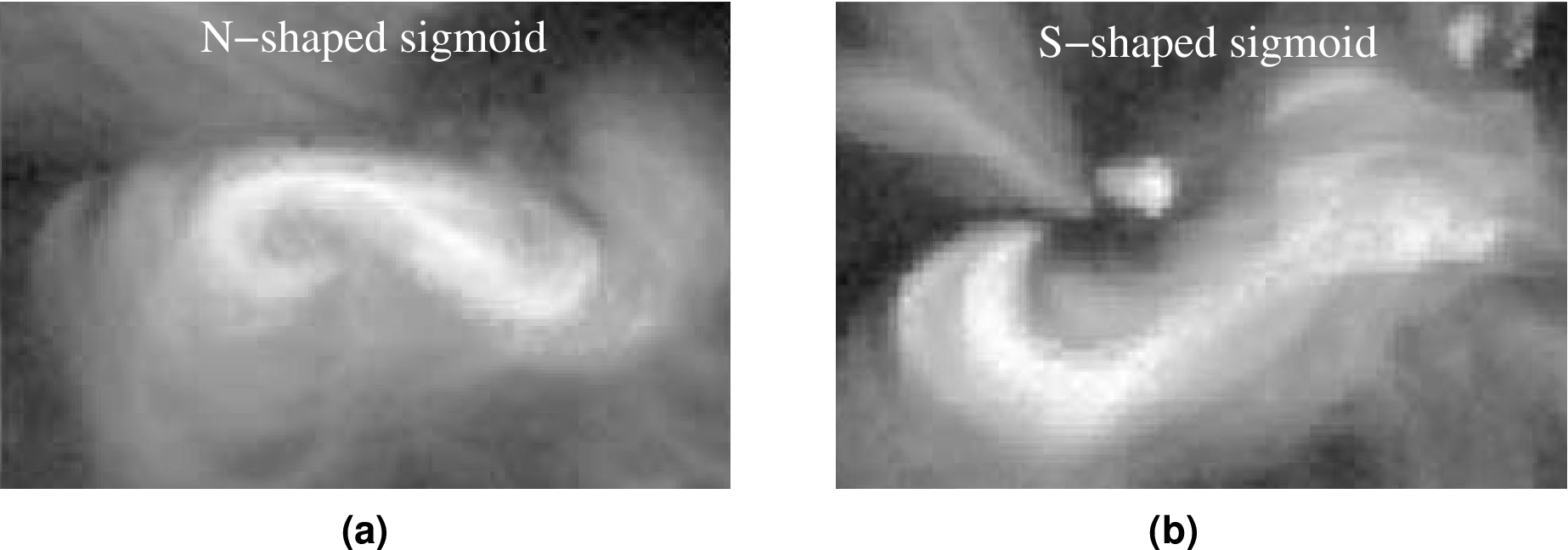}}
	\caption{Determining the helicity sign based on the shape of sigmoids above active-region filaments: N-shaped sigmoids imply negative helicity and S-shaped sigmoids imply positive helicity. Adapted from \citet{pevt02}.}\label{fig05}
\end{figure}

As illustrated in Figure \ref{fig05}, if the soft X-ray or EUV observations reveal an inverse S-shaped sigmoid above a filament, the helicity is negative (left panel of Figure \ref{fig05}); If it is an S-shape sigmoid above a filament, the helicity is positive. Since inverse S-shape looks like N-shape, inverse S-shaped sigmoids are sometimes called N-shaped sigmoids. More interestingly, the Northern hemisphere is dominated by negative helicity and Southern hemisphere is dominated by positive helicity, it means that the {\color{red}{N}}ortern hemisphere is dominated by {\color{red}{N}}-shaped sigmoids and the {\color{red}{S}}outhern hemisphere is dominated by {\color{red}{S}}-shaped sigmoids. What a coincidence!

It is noted in passing that in many cases filament spines look like either an N-shape or S-shape, corresponding to negative helicity and positive helicity, respectively. However, its universality requires further investigation since some filament channels are serpentine, and it is not unusual to find a counterexample for the correspondence.     

\subsection{Bearing sense of Filament Barbs}\label{barb}

A solar filament usually consists of a spine and several barbs. If the spine is analogous  to an expressway, barbs are like the exists of the expressway. Similar to the expressways in China and UK where the exits are right-bearing and left-bearing, respectively, filament barbs might also veer away from the filament spine rightward or leftward with an acute angle. Right-bearing and left-bearing barbs are the typical structural chirality of solar filaments.

\begin{figure} 
	\centerline{\includegraphics[width=1.0\textwidth,clip=]{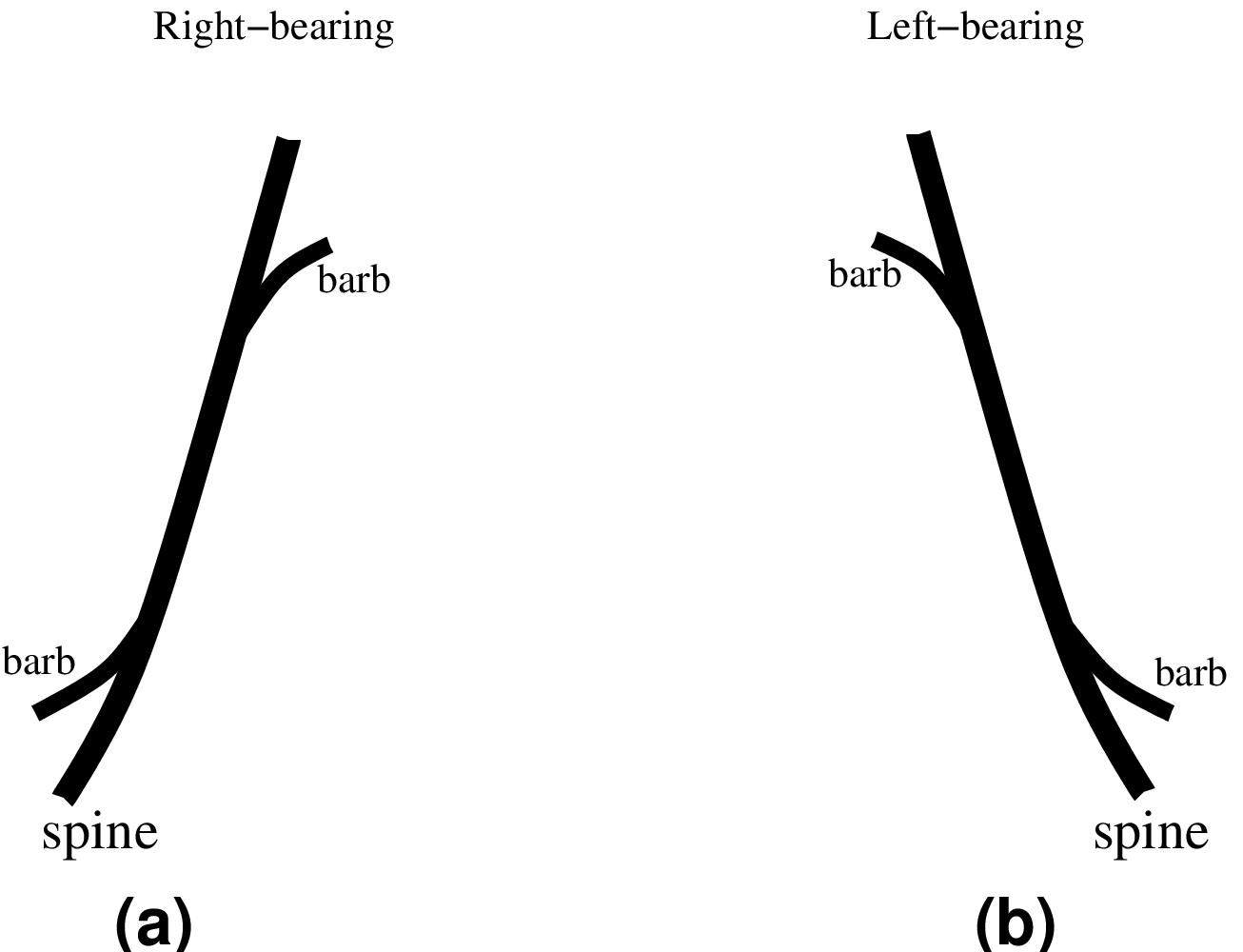}}
	\caption{Sketch showing a possible correspondence between the bearing sense of filament barbs and helicity sign proposed by \citet{mart94}, according to whom right-bearing barbs indicate negative helicity and left-bearing barbs indicate positive helicity.}\label{fig06}
\end{figure}

\citet{mart94} and \citet{mart98} proposed that there exists a one-to-one correspondence between the structural chirality and sigh of helicity. As shown in Figure \ref{fig06}, right-bearing filament barbs, which are rooted at parasitic magnetic polarity, possess negative helicity as in panel (a), whereas left-bearing filaments possess positive helicity as in panel (b).

Several research groups applied this method to a large number of solar filaments to investigate the hemispheric preference of helicity on the Sun, i.e., the majority of filaments have negative helicity in the northern hemisphere and positive helicity in the southern hemisphere. However, the results are diverse, with the strength of the preference being 58--83\% \citep{pevt03, bern05, yeat07, lim09}. 

As pointed out by \citet{chen24}, caution should be taken when using this method. First, the Sun is a sphere, so that a left-bearing barb might be misidentified as a right-bearing barb when a filament is far from the solar disk center, and vice versa. Second, there are some anomalous barbs which look like left- (or right-) bearing but composed of right- (or left-) bearing threads \citep{mart08, fili17}. Third and most importantly, it is not unusual that left-bearing barbs and right-bearing barbs might coexist in a filament \citep{pevt03b}, although the helicity is still the same between the two parts. \citet{guo10} found that Martin’s method is correct only when the filament portion is supported by a magnetic flux rope, whereas Martin’s method fails when the filament portion is supported by a sheared arcade. That is to say, this method might not be universal.

\subsection{Skew of endpoint brightenings}

\begin{figure} 
	\centerline{\includegraphics[width=1.0\textwidth,clip=]{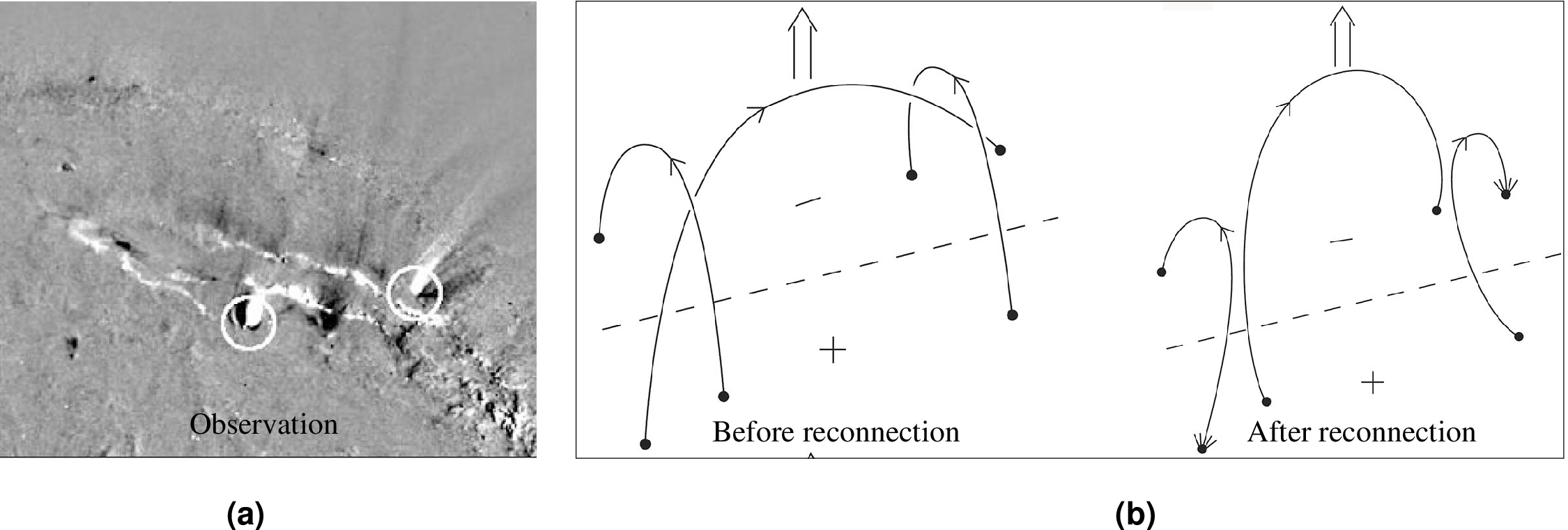}}
	\caption{Determining the axial magnetic field and helicity by the pattern of the endpoint brightenings due to interchange reconnection during filament eruption. (a) EUV image showing endpoint brightenings outside flare ribbons. (b) Interpretation of the endpoint brightenings proposed by \citet{wang09}. Adapted from \citet{wang09}.}\label{fig07}
\end{figure}

As a filament erupts, two typical features are a two-ribbon flare immediately below the erupting filament and the conjugate transient coronal holes around the two endpoints of the filament. \citet{wang09} noticed the third feature, i.e., EUV brightenings appear at the outer edge of the transient coronal holes. The endpoints brightenings are around strong network well outside the filament channel. Therefore, they proposed that, one can determine the axial magnetic direction and the helicity sign of the filament channel by combining EUV images and longitudinal magnetograms. They interpreted the endpoint brightenings to be due to interchange reconnection, as illustrated in Figure \ref{fig07}, where the axial magnetic field is from the bottom left to the upper right, hence the core magnetic field is left-handed, i.e., the helicity is negative.

Again, as mentioned above, the role of longitudinal magnetograms is for the purpose to determine the direction of the axial magnetic field. When we simply want to determine the helicity sign, longitudinal magnetograms are not needed.

\begin{figure} 
	\centerline{\includegraphics[width=1.0\textwidth,clip=]{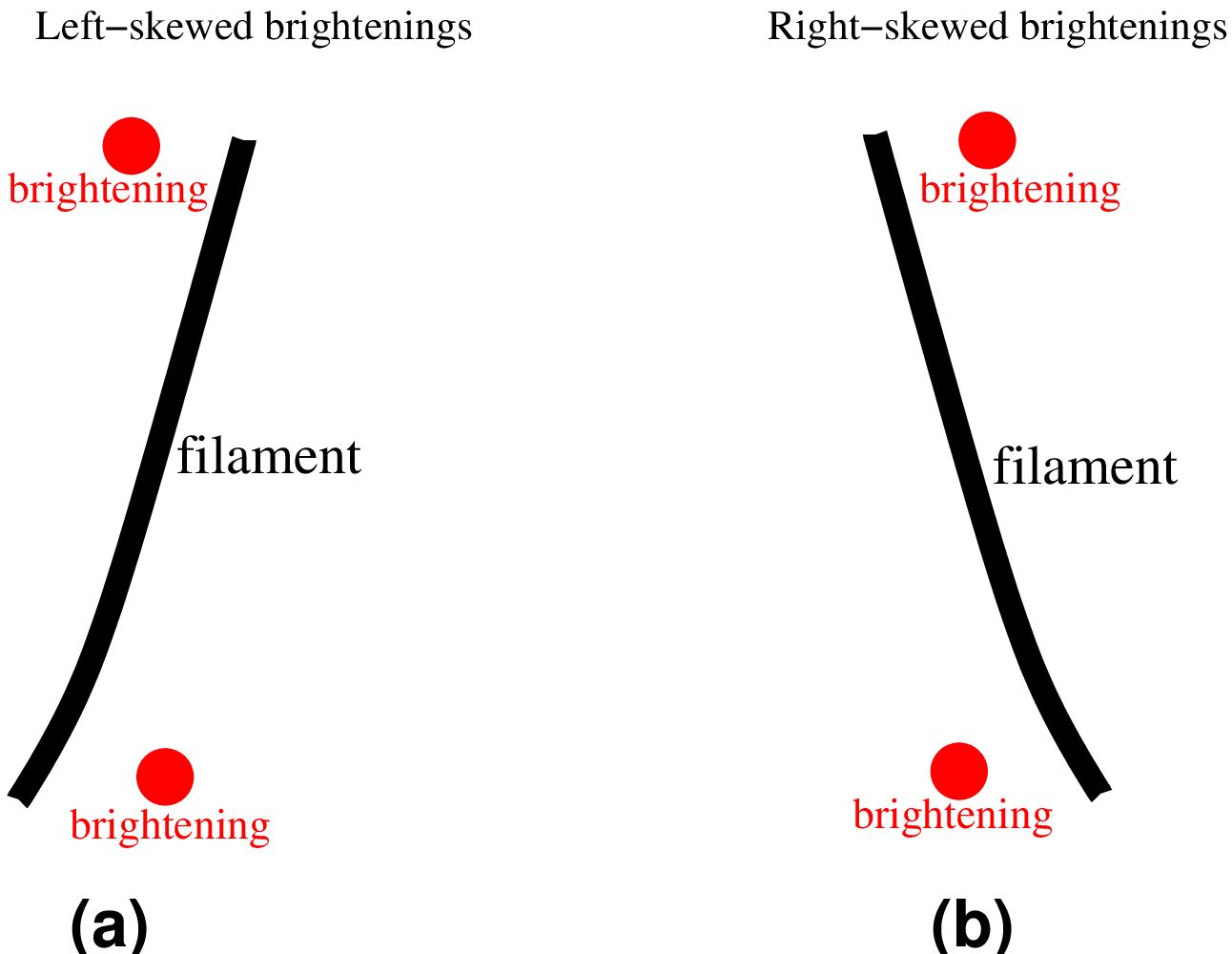}}
	\caption{Sketch showing how to determine the helicity sign based on the skew of the conjugate drainage sites relative to the preeruptive filament as proposed by \citet{chen14}. (a) Left-skewed drainage sites imply negative helicity. (b) Right-skewed drainage sites imply positive helicity.}\label{fig08}
\end{figure}

While the endpoint brightenings might sometimes be due to interchange reconnection between the core magnetic field and the envelope field of the filament channel \citep{wang09}, \citet{chen14} found another mechanism for the conjugate endpoint brightenings, i.e., filament drainage, which might be more often. During each filament eruption, only some of the cold materials are pulled out to form the core of the resulting CME, other cold materials drain down along the two legs of the magnetic field lines. When these plasmas bombard the two footpoints, conjugate brightenings are formed. \citet{chen14} noticed that when a filament with negative helicity erupts, the conjugate brightenings are left skewed from the filament spine no matter the supporting magnetic field is a flux rope or a sheared arcade (Figure \ref{fig08}a); When a filament with positive helicity erupts, the conjugate brightenings are right skewed no matter the supporting magnetic field is a flux rope or a sheared arcade (Figure \ref{fig08}b). Hence, they proposed that, as illustrated in Figure \ref{fig08}, if the conjugate brightenings are left skewed relative to the initial filament, the helicity of the filament channel should be negative; If the conjugate brightening are right skewed with respect to the preeruptive filament, the helicity is positive.

\citet{ouyang17} applied this method to 571 solar filaments, and found the strongest hemispheric preference of helicity, i.e., 92.6\% of the solar filaments in solar cycle 24 have negative helicity in the northern hemispheric and positive helicity in the southern hemisphere.

\section{Magnetic configuration}\label{s:3}
\subsection{Curvature Radius of Dipped Magnetic Field Lines}\label{s:31} 

As the plasma density of solar filaments is two orders of magnitude larger than that of the ambient corona, such heavy materials should be supported against gravity by upward Lorentz force of coronal magnetic field. To keep the cold dense plasmas in stable equilibrium during the lifetime of filaments, concave-upward, i.e., dipped, magnetic field is the best choice. Therefore, it is believed that a majority of solar filaments, especially those relatively static filaments, are supported by dipped magnetic field \citep{kupe74, parenti14}. One important quantity of the dipped magnetic configuration is the curvature radius of the concave-upward magnetic field lines.

In addition to the transverse oscillations which have been investigated since 1960s, \citet{jing03} found that a solar filament may experience longitudinal oscillations upon perturbations from a nearby microflare. In this kind of oscillations, filament threads move along the long axial direction of filament threads. Note that filament threads are believed to trace local magnetic field \citep{liny05, hana17}. It should be pointed out that filament longitudinal oscillations should not be misunderstood as oscillations along the filament spine, i.e., lateral displacement of filaments is not unique for filament transverse oscillations \citep{pant16, chenjl17}. Longitudinal oscillations are also accompanied by lateral displacement of filaments as filament threads, and hence the magnetic field lines, deviate from the filament spine direction by $10^\circ$--$30^\circ$ \citep{hana17}.

Since filament longitudinal oscillations are along the local magnetic field, generally speaking Lorentz force does not play a role, and gravity is the main restoring force. In this circumstance, the longitudinal oscillation of filament threads can be described by the pendulum model \citep{luna12, zhang12}. In this model, the oscillation period ($P$) is determined by the curvature radius of the dipped magnetic field ($R$), i.e., 

\begin{equation}
	P=2\pi\sqrt{R/g} \label{eq1}
\end{equation}

\noindent
where $g$ is the gravitational acceleration near the Sun.

By analyzing the high-resolution images obtained with the Hinode mission, \citet{zhang12} were able to trace the trajectory of an oscillating prominence, hence to measure the curvature radius of the supporting magneti field line. The observed oscillation period is consistent with the value derived from Equation (\ref{eq1}), verifying the validity of the pendulum model. This implies that we can infer the curvature radius of the supporting magnetic field lines based on the observed period of longitudinal oscillations of solar filaments. It is noted that the variation of the flux tube cross-section does not significantly influence the oscillation period \citep{luna16}.

Such a pendulum model was later applied to derive the curvature radius of supporting magnetic field of solar filaments \citep[e.g.,][]{pant16, tan23}. However, it should be pointed out that there are three caveats about the application of this method. First, when the magnetic field is too weak, e.g., the plasma $\delta$ is close or larger than unity \citep{zhou18}, the longitudinally moving filament threads would deform the magnetic configuration \citep{lit12, zhangly19}, which would prolong the oscillation period compared to the pendulum model. Second, filament longitudinal oscillations generally damp out within 3 periods \citep{luna18}. However, some filaments might experience dampless oscillations, in particular before they erupt \citep{chen08, lit12}. \citet{niyw22} found that if the decayless longitudinal oscillations are driven by periodic jets, the averaged oscillation period depends on both the intrinsic period, i.e., the one determined by the magnetic curvature radius, and the external driving period of the jets. With numerical simulations, they derived a formula to relate the observed periods of filament oscillations and the intermittent jets to the intrinsic period of the filament, by which we can more precisely determine the magnetic curvature radius with Equation (\ref{eq1}). Third, if a flux tube is twisted by more than 2 turns with two threads residing at the two dips, there is interference between the longitudinal oscillations of the two magnetically connected threads, which results in $\sim$10\% deviation from the intrinsic period \citep{zhou17}. 

It is noted that longitudinal and transverse oscillation might coexist in a filament \citep{liak20, daij23, tan23}, which provides a unique opportunity to derive both the strength and the curvature radius of the supporting magnetic field of the filament.

\subsection{Sheared arcade versus flux rope}
Since filaments correspond to dense plasmas rather stably suspended in the much tenuous corona, it is generally believed that they are supported against gravity by magnetic dips. The existence of magnetic dips is implied also by the ubiquitous longitudinal oscillations of filament threads.

\begin{figure} 
	\centerline{\includegraphics[width=1.0\textwidth,clip=]{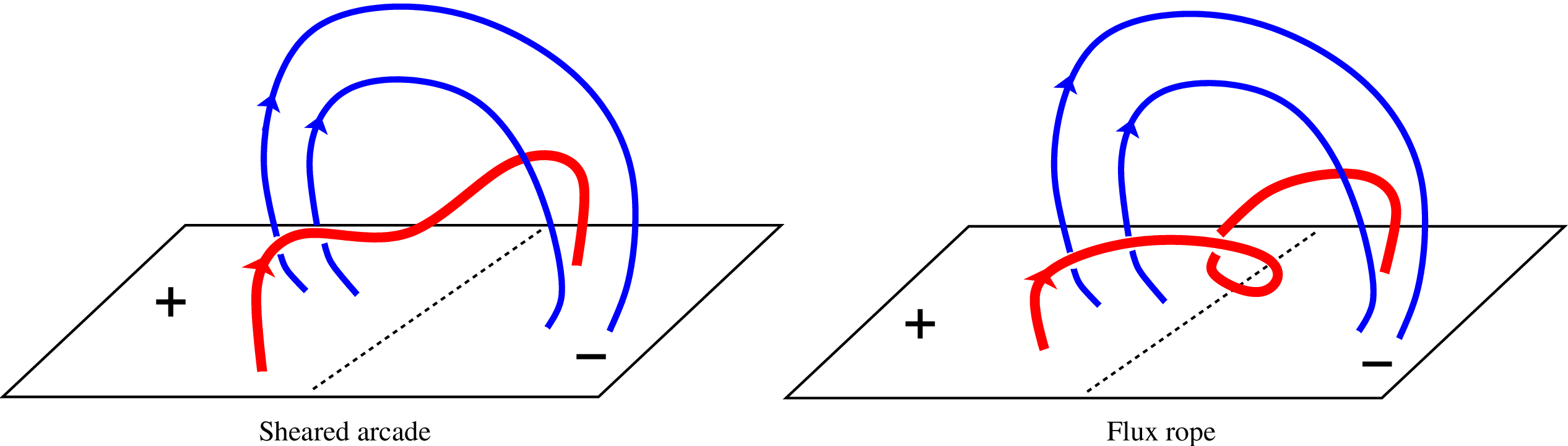}}
	\caption{Sketch showing the two typical magnetic configurations with dips, i.e.,  sheared arcade and flux rope. The red lines are the core magnetic field where magnetic dips support cold dense materials of the filament, and the blue lines are the envelope magnetic field. The core field lines in the flux rope are twisted by one or more turns.}\label{fig09}
\end{figure}

Historically, two types of magnetic configurations with magnetic dips have been proposed, i.e., dipped arcades described by the KS model \citep{kipp57} and flux ropes described by the KR model \citep{kupe74}, which correspond to the normal-polarity filaments and inverse-polarity filaments, respectively. Realizing that the axial magnetic field is the main component of the magnetic field for filaments \citep{rust94}, the KS model was later generalized to the sheared arcade model \citep{aula02} as illustrated by the left panel of Figure \ref{fig09}, and the flux rope model was also extended from 2-dimensional a magnetic island to a 3D flux rope \citep{vanb89, aula06} as illustrated by the right panel of Figure \ref{fig09}. It is noted that some filaments may be simply a collection of cold chromospheric plasmas being injected into the corona from one polarity and being streaming down to the chromosphere at the other polarity \citep{wang99, zou16}. In this case, magnetic dips are not necessary \citep{karp01}.

From 1960s to 1980s, there were continued efforts to measure the magnetic field of solar filaments, which verified the existence of normal-polarity and inverse-polarity filaments \citep{lero84, bomm94}. Such measurements became scarce since 1990s. Coronal magnetic extrapolation has remained to be the only effective way to determine the supporting magnetic configuration of filaments for a long time. All these led to the long debate whether flux ropes are always the preeruptive structure of CMEs \citep{ouyang15}. Therefore, developing other independent methods to deduce the magnetic configuration has been a pressing necessity before routine magnetic measurements of solar filaments are available.

\begin{figure} 
	\centerline{\includegraphics[width=0.8\textwidth,clip=]{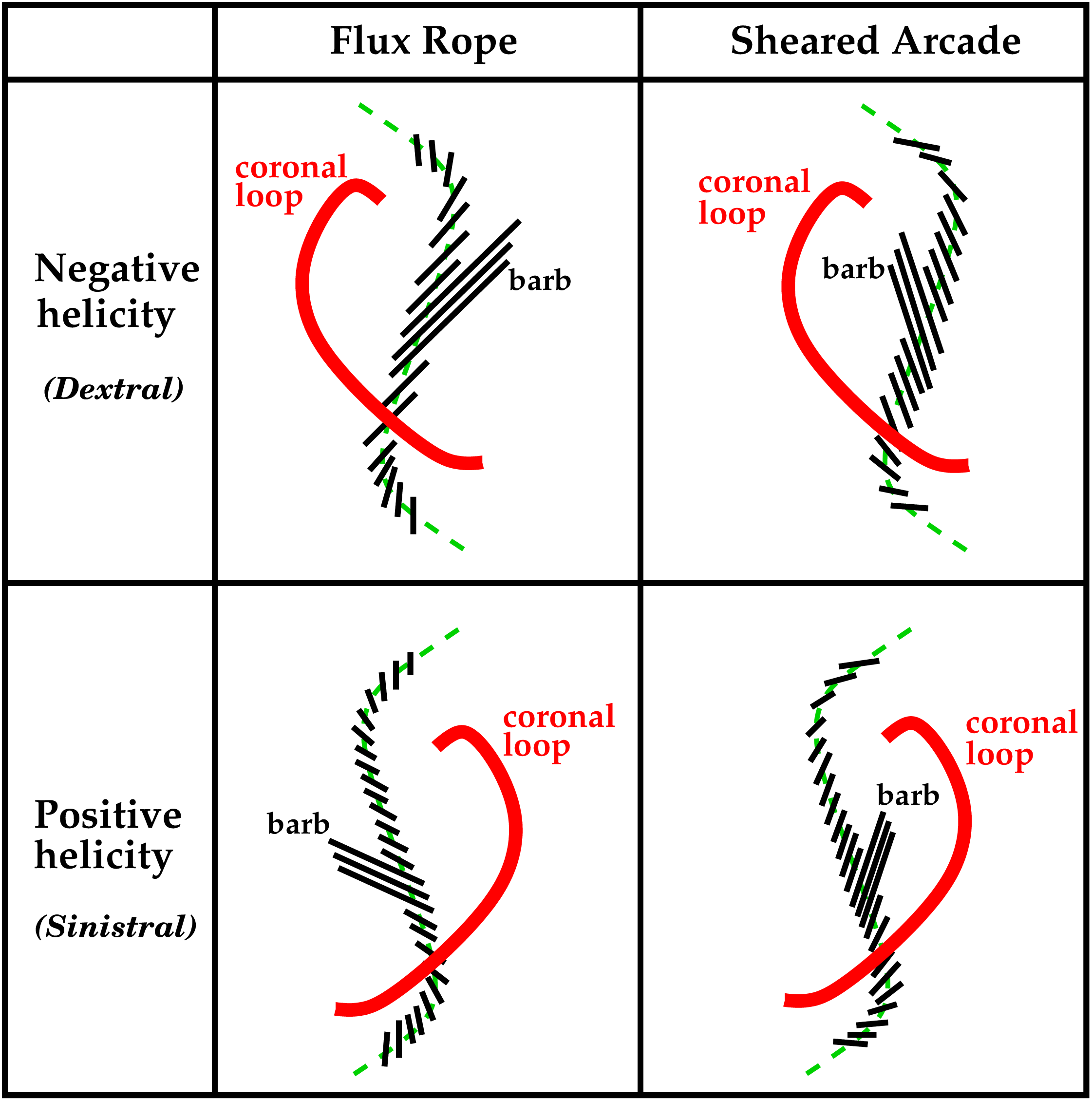}}
	\caption{Sketch showing the correspondence between the magnetic helicity
		and the filament patterns in two magnetic configurations, i.e., flux ropes
		and sheared arcades. From \citet{chen20}.}\label{fig10}
\end{figure}

As we mentioned above, \citet{mart94} proposed that filaments with right-bearing/left-bearing barbs have negative/positive helicity, which we call Martin's rule. Note that a filament might have both left-bearing and right-bearing barbs \citep{pevt03b}, hence the bearing sense was generally judged based on the dominant one. However, \citet{guo10} found that this rule is not universal, i.e., for a filament with both right-bearing barbs and left-bearing barbs, its helicity was found to be negative consistently. With magnetic extrapolation, they found that the filament portion with right-bearing barbs is supported by a flux rope and the portion with left-bearing barbs is supported by a sheared arcade. \citet{chen14} generalized this result and proposed an indirect method to deduce whether a filament is supported by a flux rope or a sheared arcade: If the relationship between the helicity sign and the filament barbs bearing sense follows Martin’s rule, the filament is supported by a flux rope; If the relationship is against Martin’s rule, the filament is supported by a sheared arcade. It means that famous pattern about the relationship between helicity and filament barbs \citep[Figure 10 in][]{mart98} might be applicable only for the filaments supported by flux ropes. For a more general application, \citet{chen20} proposed a modified pattern, which is illustrated in Figure \ref{fig10}.

As explained in \citet{chen14}, the procedure consists of two steps: (1) To determine the helicity sign of a filament: This can be done with any of the methods described in Sect. \ref{s:2} (vector magnetograms can do the job as well); (2) To determine the bearing sense of filament barbs or filament threads. Once the two steps are done, we can use Figure \ref{fig10} to deduce whether the filament is supported by a flux rope or sheared arcade.
\citet{ouyang15} applied this method to two filament eruption events, and found one is supported by a flux rope and the other by a sheared arcade. With this, they concluded that flux rope is not a necessary condition for the progenitor of CMEs. \citet{ouyang17} further applied this method to 571 filament eruption events observed from 2010 May to 2015 December and revealed that 89\% of the filaments are supported by the flux rope magnetic configuration and 11\% of the filaments are supported by the sheared arcade configuration.

\section{Twist Strength of Flux Ropes}\label{s:4}

Although not every filament is supported by a flux rope, the flux rope configuration has received continued attention. First, as revealed by \citet{ouyang17}, a majority of filaments, in particular the quiescent filaments, are supported by flux ropes. Second, several ideal MHD triggering mechanisms for filament eruptions are associated with flux ropes, such as kink instability and torus instability \citep{chen11}. Third, surrounding and below a flux rope readily lies a magnetic QSL or even separatrix (In contrast, there would be a QSL surrounding a sheared arcade only when the arcade is strongly sheared or after magnetic reconnection). Hence, once the flux rope is triggered to rise, it is natural to form a current sheet below the flux rope, inside which the reconnection would naturally lead to the filament eruption and a CME.

\begin{figure} 
	\centerline{\includegraphics[width=0.8\textwidth,clip=]{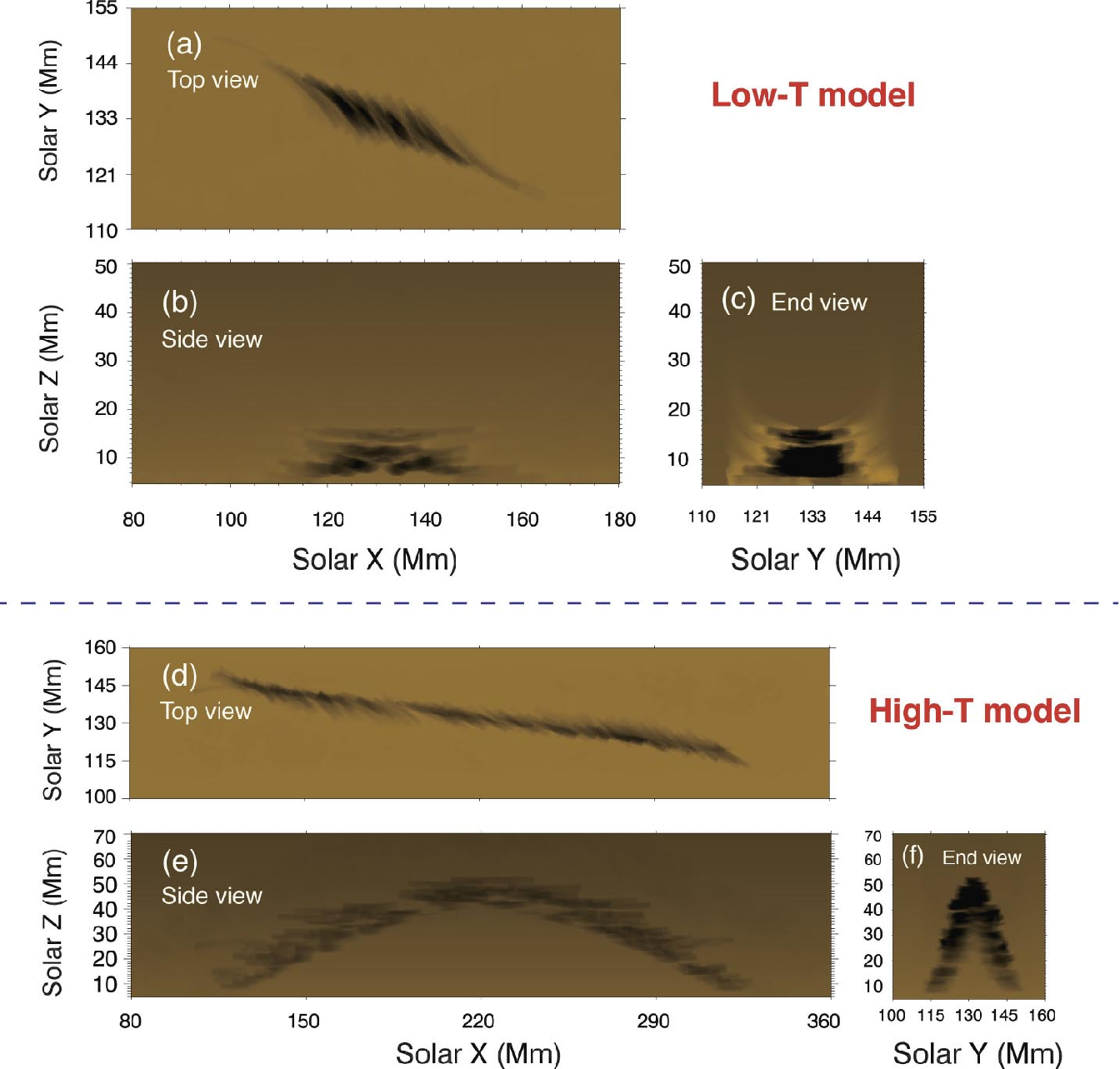}}
	\caption{Different appearance from various views of two filaments with low and high twists, respectively, based on MHD forward modeling. From \citet{guojh22}.}\label{fig11}
\end{figure}

One key parameter of flux ropes is the twist number, i.e., how many turns are the magnetic field lines twisted? While the exact number of magnetic twist can be obtained only when we know the real 3D distribution of the magnetic field in the corona, \citet{guojh22} tried to explore whether we can distinguish between strong and weak twists based on EUV imaging observations of solar filaments. The line of thought is straightforward: To conduct forward modeling of two filaments formed in flux ropes with weak and strong twists, and then compare the difference in synthesized EUV images. They selected two flux ropes, one with a twist of 1.07 and the other with a twist of 2.14, their pseudo-3D simulations were performed, and the numerical results were used to synthesize EUV images. The different images from three viewing angles are displayed in Figure \ref{fig11}, which indicates that (1) weakly-twisted flux ropes tend to have longer threads in the top view; and (2) weakly-twisted flux ropes tend to have more evident horns in the end view. These features, once confirmed, can be applied to determine the twist strength of flux rope qualitatively or even semi-quantitatively.

\section{Final Remarks}\label{s:5}

In the ancient times when there were no advanced diagnostic tools as in modern medicine, frequently doctors around world diagnosed diseases based on the observations of clinical symptoms. For example, in the traditional Chinese medicine, doctors diagnose diseases based on observation, auscultation and olfaction, interrogation, pulse taking and palpation, among which observing the appearance of the patient is one important method since different diseases may manifest differently on our face and skin. Doctors have mastered the empirical relationship between diseases and skin appearance. With the same philosophy, the solar community has developed many indirect methods to diagnose magnetic parameters of solar filaments based on imaging observations. In this review paper, we summarized various efforts on diagnosing several magnetic parameters of filament channels based on the imaging observations of solar filaments, including the helicity sign, the curvature radius of magnetic field lines, the magnetic configuration, and even the twist strength of flux ropes. Such a methodology, as we call solar filament physiognomy, provides an effective approach to derive various parameters easily based on their EUV and H$\alpha$ imaging observations. As we have witnessed,  this phenomenological approach can offer some key information of the magnetic field that might not be accurately obtained from vector magnetic measurements.

Filaments composed by cold plasmas remain a hot topic in solar physics. First, their thermal structure, including density and temperature distributions, is still unclear \citep[e.g.,][]{guna18}. The thermal distribution is never trivial as it provides stringent constraint for the coronal heating function, which is a big question in solar physics. Second, their magnetic distribution with singular topology is crucial for understanding the breakdown of flux conservation in the low-resistivity limit \citep{low15}, hence for understanding the triggering and driving of solar eruptions at large. With the advent of large telescopes, such as the 4-meter DKIST \citep{woge21} and the 2.5-meter WeHost \citep{fang19}, more and more fine structures of solar filaments will be disclosed. It is expected that more physics can be revealed as we believe that every detail happens for a reason.

%

%

%
\begin{acks}
The author thanks the anonymous referee for the constructive suggestions. 
\end{acks}

%
%
\begin{fundinginformation}
The research was supported by the National Key Research and Development Program of China (2020YFC2201200) and NSFC (12127901).
\end{fundinginformation}

%
%
%
%

%
%
%
%

%

\end{document}